\documentclass{ws-procs9x6}

\begin{document}
\title{ MASSIVE HYBRID STARS WITH STRANGENESS}

\author{ T. TAKATSUKA$^*$ and T. HATSUDA}

\address{ Theoretical Research Division, Nishina Center, RIKEN, 
Wako 351-0198, Japan\\
$^*$E-mail: takatuka@iwate-u.ac.jp}

\author{ K. MASUDA}
\address{Department of Physics, The University of Tokyo, 
Tokyo 113-0033, Japan}

\begin{abstract}
How massive the hybrid stars could be is discussed by 
a \lq\lq 3-window model\rq\rq proposed from 
a new strategy to construct the equation of state 
with hadron-quark transition.  
It is found that hybrid stars have a strong potentiality 
to generate a large mass compatible with two-solar-mass 
neutron star observations.
\end{abstract}

%\keywords{Style file; \LaTeX; Proceedings; World Scientific Publishing.}

\bodymatter

\section{ Introduction}\label{aba:sec1}
It seems a recent consensus that hyperons (Y) are sure to participate 
in neutron star (NS) cores, increasing the population with 
the increase of baryon density ($\rho$).\cite{taka04}
The Y-mixing, as a manifestation of strangeness degrees of freedom, 
plays a dramatic role in NS properties, that is, 
it causes an extreme softening of the EOS
\cite{nishi01}$^-$\cite{dapo10}, 
leading to the problem that the maximum mass 
($M_{max}$) of NSs cannot exceed even the 
1.44 $M_{\odot}$ observed for PSR1913+16.  
This conflict between the theory and the observation 
becomes more serious by a very recent finding of 
2$M_{\odot}$-NSs\cite{demo10}$^,$\cite{anto13}.  
In a pure hadronic framework, it has been pointed out 
that the introduction of a \lq\lq universal 3-body force\rq\rq 
acting on all the baryons BBB 
(i.e., not only on NNN but also on NNY, NYY and YYY) 
is a promising candidate to solve the problem.\cite{taka08}

The aim of this paper is to discuss another solution for 
the problem by extending the framework from pure hadron to hadron 
(H) plus quark (Q) degrees of freedom.  
We address how the hybrid stars with 
H-Q transition core could be massive, 
by a new approach not restricted to the conventional 
Gibbs or Maxwell condition.  
Our new strategy is to divide the equation of state 
(EOS) into 3 density regions, i.e., pure 
H-EOS for $\rho\leq\rho_H$, HQ-EOS for 
$\rho_H\leq\rho\leq\rho_Q$ and pure Q-EOS for 
$\rho\geq\rho_Q$, characterized as 
\lq\lq 3-window model\rq\rq\cite{taka12}. 
The motivation comes from the considerations: 
(i) Pure hadronic EOS gets uncertain with increasing 
$\rho$ because of finite size hadrons composed of 
quarks and gluons.  
(ii) Pure quark matter EOS becomes unreliable with decreasing 
$\rho$
due to the deconfined-confined transition.  
(iii) Therefore, to discuss the H-Q transition by 
extrapolating the pure H-EOS from a lower 
density side and the pure Q-EOS from a higher 
density side is not necessarily justified.  
Our basic idea is to supplement the very 
poorly known HQ-EOS by sandwitching it in between the relatively 
certain H-EOS and Q-EOS, and construct the HQ-EOS 
by a phenomenological interpolation.

\begin{figure}
\psfig{file=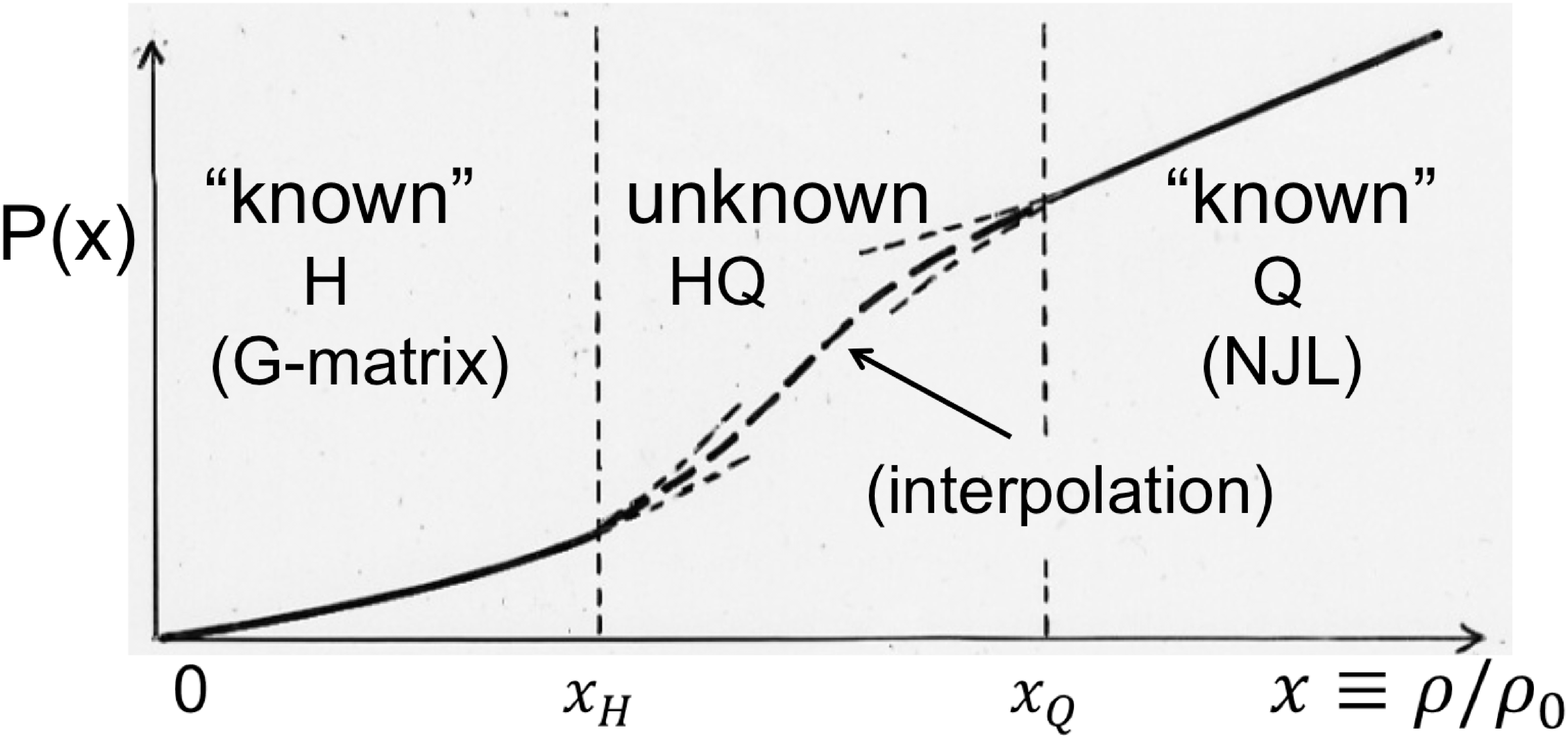,width=4.2in}
\caption{Schematic illustration of 
\lq\lq 3-windouw model\rq\rq.  
A very poorly known HQ-EOS is interpolated by sandwiching 
it in between a H-EOS and a Q-EOS relatively known.
}
\label{aba:fig1}
\end{figure}

\section{Approach}\label{aba:sec2}
According to the \lq\lq 3-window model\rq\rq, 
we construct the EOS with H-Q transition.  
In our preceding works\cite{masu13a}$^,$\cite{masu13b}, we have tried this line 
of approach from a view of a smooth crossover for the 
H-Q transition region and found that the hybrid stars 
satisfy $M_{max}\geq2M_{\odot}$.  
There the pressure $P(\rho$) was interpolated as 
$P(\rho)=P_H(\rho)f_-(\rho)+P_Q(\rho)f_+(\rho)$ by a 
$\rho$-dependent weight function 
$f_{\pm}(\rho)=(1\pm$ tanh $[(\rho-\bar\rho)/\Gamma])$ with parameters $\bar\rho$ and $\Gamma$, in an analogy to very hot 
QCD transition.  
Due to $f_{\pm}(\rho)$, however, the interpolated 
HQ-EOS approaches only asymptotically to H-EOS with 
decreasing $\rho$ (Q-EOS with increasing $\rho$).  
But such a way of interpolation is not unique.  
As a complementary work, here we try more general interpolation 
and make exact matching at discrete boundaries 
($\rho_H$ and $\rho_Q$).  

As in the preceding work\cite{masu13b}, we take the H-EOS with Y
(denoted by TNI2) from a G-matrix effective interaction approach.  
The TNI2 H-EOS satisfies the saturation property of 
symmetric nuclear matter and has an incompressibility 
$\kappa=250$MeV consistent with experiments.  
We use the 3-flavor Q-EOS from the NJL model including 
a repulsive effect from vector interaction with 
the strength $g_v=(0-1.5)G_s$ 
($G_s$ being the strength of scalar interaction).  
As an interpolation function, we take 
$P_{HQ}(x)=ax^m+bx^n+c$ with $x\equiv\rho/\rho_0$ 
($\rho_0=0.17$/fm$^3$ being the nuclear density).  
Then, the energy density $\epsilon$ is obtained from 
$P=\rho^2\partial(\epsilon/\rho)/\partial\rho$ as 
$\epsilon_{HQ}(x)=(a/(m-1))x^m+(b/(n-1))x^n+dx-c$.  
Four coeficients \{a, b, c, d\} are determined for a given set 
\{m, n\} and \{$x_H\equiv\rho_H/\rho_0$, $x_Q\equiv\rho_Q/\rho_0$\} 
by a matching of $P$ and $\epsilon$ at phase boundaries.  
By running the set of \{m, n\} and \{$x_H$, $x_Q$\}, 
the solution is searched under the conditions; 
(i) $P(x)>0$ and $\partial P/\partial x\geq0$ 
(thermodynamic stability), (ii) $v_s/c=(\partial P/\partial\epsilon)^{1/2}\leq1$ 
(sound velocity less than light velocity), (iii) 
$x_H>1$ (no experimental evidence for quark degrees of freedom at 
$\rho\leq\rho_0$). 

\begin{table}[b]
\tbl{Some results for NS models}
{\begin{tabular}{@{}cccccccccc@{}}\toprule
CASE & $x_H$ & $x_S$ & H-EOS & Q-EOS & m & n 
& $M_{max}/M_{\odot}$ & $R/k_m$ & $\rho_c/\rho_0$ \\\colrule
1 & 1.5\hphantom{00} & \hphantom{0}5.5 & \hphantom{0}TNI2 
& \hphantom{0}$g_v=0.5G_s$ & \hphantom{0}0.2 & \hphantom{0}-2.6 
& \hphantom{0}2.61 & \hphantom{0}13.38 & \hphantom{0}3.99 \\
2 & 1.5\hphantom{00}&\hphantom{0}6.0&\hphantom{0}TNI2
& \hphantom{0}$g_v=0.5G_s$&\hphantom{0}0.2&\hphantom{0}-2.6&\hphantom{0}2.59 
&\hphantom{0}13.27&\hphantom{0}3.90 \\
3 & 1.5\hphantom{00}&\hphantom{0}7.0&\hphantom{0}TNI2 
&\hphantom{0}$g_v=0.5G_s$&\hphantom{0}0.2&\hphantom{0}-2.6 
&\hphantom{0}2.53&\hphantom{0}12.08&\hphantom{0}4.52 \\
4 & 1.5\hphantom{00}&\hphantom{0}8.0&\hphantom{0}TNI2
& \hphantom{0}$g_v=0.5G_s$&\hphantom{0}0.2&\hphantom{0}-2.6 
&\hphantom{0}2.48&\hphantom{0}12.56&\hphantom{0}4.35 \\
5 & 1.5\hphantom{00}&\hphantom{0}7.0&\hphantom{0}TNI2 
&\hphantom{0}$g_v=1.5G_s$&\hphantom{0}0.2&\hphantom{0}-2.6
&\hphantom{0}3.08&\hphantom{0}13.73&\hphantom{0}3.34 \\
6 & 1.5\hphantom{00}&\hphantom{0}7.0&\hphantom{0}TNI2 
&\hphantom{0}$g_v=1.0G_s$&\hphantom{0}0.2&\hphantom{0}-2.6 
&\hphantom{0}2.86&\hphantom{0}13.28&\hphantom{0}3.94 \\
7 & 1.5\hphantom{00}&\hphantom{0}7.0&\hphantom{0}TNI2
& \hphantom{0}$g_v=0$&\hphantom{0}0.2&\hphantom{0}-2.6 
&\hphantom{0}1.99&\hphantom{0}12.30&\hphantom{0}4.85 \\
8 & 1.5\hphantom{00}&\hphantom{0}7.0&\hphantom{0}TNI2 
&\hphantom{0}$g_v=0.5G_s$&\hphantom{0}2.6&\hphantom{0}-0.2
&\hphantom{0}2.62&\hphantom{0}13.44&\hphantom{0}4.05 \\
9 & 1.5\hphantom{00}&\hphantom{0}7.0&\hphantom{0}TNI2 
&\hphantom{0}$g_v=0.5G_s$&\hphantom{0}1.2&\hphantom{0}-1.2 
&\hphantom{0}2.61 & \hphantom{0}13.44&\hphantom{0}3.73 \\\botrule
\end{tabular}}
\end{table}

\section{Some Results and Remarks}\label{aba:sec3}
Some examples for numerical results are shown in Table 1.  
We note the following points: (i) Within the present 
interpolation function, we have several hybrid stars with $M_{max}\simeq(2-3)M_{\odot}$.  
It can be as massive as $3M_{\odot}$-NSs.  (ii) 
The dependence of $M_{max}$ on \{m n\} and \{$x_H$, $x_Q$\} is rather small: 
For a fixed \{m=0.2, n=-2.6, $x_H=1.5$\} and $g_v=0.5G_s$, 
$M_{max}$ changes slightly, $(2.61\rightarrow2.48)M_{\odot}$ 
according to 
$x_Q=(5.5\rightarrow8.0)$.  
For a fixed \{$x_H=1.5$, $x_Q=0.7$\} and $g_v=0.5G_s$, 
the functional dependence of $M_{max}$ is also small as 
$M_{max}$=(2.53, 2.62, 2.61)$M_{\odot}$ for 
(m, n)=(0.2, -2.6), (2.6, -0.2), (1.2, -1.2). (iii) 
The $g_v$-dependence of $M_{max}$ is remarkable as 
$M_{max}$=(1.99, 2.53, 2.86, 3.08)$M_{\odot}$ 
according to $g_v$=(0.0, 0.5, 1.0, 1.5)$G_s$
(CASE 7, 3, 6, 5).  (iv) 
Since $x_Q>\rho_c$ (the central density), 
our hybrid stars do not have pure Q-matter core  but 
H-Q transient core.  
In the calculations we have found that the $x_H$ as lower as 
(1.5-2.5) is necessary for the solution to exist.  
This may suggest a picture that the Q-degrees of freedom begins 
to work at rather low density as has been discussed 
from a view of quark percolation in nuclear medium\cite{baym79}.  

To summarize, our hybrid stars from the 
\lq\lq 3-window model\rq\rq can generate the $M_{max}$ compatible with 
2$M_{\odot}$-NS observations, as far as the 
Q-degrees of freedowm sets on from a rather low density 
($\sim1.5\rho_0$) and the Q-EOS is stiff enough.  
The present work supports the results in our preceding papers.  
Finally, we want to stress that the quark 
degrees of freedom in NS cores 
has a potentiality enough to account for 
the existence of $2M_{\odot}$-NS.

\section*{Acknowledgements}
This research was supported by 
JSPS Grant-in-Aid for Scientific Research 
(B) No.22340052 and by RIKEN 2012 
Strategic Program for R\&D.

\bibliographystyle{ws-procs9x6}
\bibliography{ws-pro-sample}

\end{document}